\documentclass[onecolumn]{aa}
\usepackage{graphicx}
\usepackage{txfonts}
\usepackage{natbib}
\usepackage{graphicx}

\bibpunct{(}{)}{;}{a}{}{,} 

\begin{document}

\title{Roche lobe effects on the atmospheric loss of \\
``Hot Jupiters''}

   \author{N. V. Erkaev
          \inst{1}
          \and
          H. Lammer
          \inst{2}
          \and
          Yu. N. Kulikov
          \inst{3}
          \and
          D. Langmayr
          \inst{2}
          \and
          F. Selsis
          \inst{4}
          \and
          G. F. Jaritz
          \inst{5}
          \and
          H. K. Biernat
          \inst{2},\inst{5}}


   \institute{Institute for Computational Modelling, Russian Academy of Sciences, Krasnoyarsk 36, Russian Federation\\
                \email{erkaev@icm.krasn.ru}
         \and
                Space Research Institute, Austrian Academy of Sciences, Schmiedlstrasse 6, A-8042 Graz, Austria\\
                \email{helfried.biernat@oeaw.ac.at, helmut.lammer@oeaw.ac.at, daniel.langmayr@oeaw.ac.at}
         \and
                Polar Geophysical Institute, Russian Academy of Sciences, Khalturina 15, 183010 Murmansk, Russian Federation\\
                \email{kulikov@pgi.ru}
         \and
                Centre de Recherche en Astrophysique de Lyon (CRAL), and Ecole Normale Sup\'{e}rieure (ENS), Lyon,
                France\\
                \email{franck.selsis@ens-lyon.fr}
         \and
                Institute for Physics, University of Graz, Universit\"atsplatz 5, A-8010 Graz, Austria\\
                \email{gerald.jaritz@stud.uni-graz.at}}


\abstract{Observational evidence of a hydrodynamically evaporating
upper atmosphere of HD209458b (Vidal-Madjar et al. 2003; 2004) and
recent theoretical studies on evaporation scenarios of ``Hot
Jupiters'' in orbits around solar-like stars with the age of the Sun
indicate that the upper atmospheres of short-periodic exoplanets
experience hydrodynamic blow-off conditions resulting in loss rates
of the order of about 10$^{10}$--10$^{12}$ g s$^{-1}$ (Lammer et al.
2003; Yelle 2004; Baraffe et al. 2004; Lecavlier des Etangs et al.
2004; Jaritz et al. 2005, Tian et al. 2005; Penz et al. 2007). By
studying the effect of the Roche lobe on the atmospheric loss from
short-periodic gas giants we found, that the effect of the Roche
lobe can enhance the hydrodynamic evaporation from HD209458b by
about 2 and from OGLE-TR-56b by about 2.5 times. For similar
exoplanets which are closer to their host star than OGLE-TR-56b, the
enhancement of the mass loss can be even larger. Moreover, we show
that the effect of the Roche lobe raises the possibility that ``Hot
Jupiters'' can reach blow-off conditions at temperatures which are
less than expected ($< 10000$ K) due to the stellar X-ray and EUV
(XUV) heating.

\keywords{exoplanets -- Roche lobe -- atmospheric loss}}

\maketitle

\section{Introduction}

Since the observation of an extended and evaporating upper
atmosphere around the Jovian-like exoplanet HD209458b by
\citet{Vidal-Madjar2003}, different blow-off scenarios for
evaporating hydrogen-rich atmospheres are discussed in the
literature \citep{Sasselov2003,Lammer2003,Lecavelier2004,Yelle2004,
Vidal-Madjar2004,Griessmeier04,Baraffe2004,Erkaev2005,Tian2005,Penz2006}.

Because hydrogen-rich upper atmospheres of short-periodic gas giants in orbits
with semi-major axes less than $<0.1$ AU can be heated to temperatures of up to
10000--20000 K by X-rays and EUV (XUV) radiation of the central star (Lammer et
al. 2003; Yelle 2004), they experience hydrodynamic blow-off. \citet{Yelle2004}
studied in detail the photochemistry of the thermospheres of ``Hot Jupiters''
and found that the lower thermosphere may be cooled primarily by infra red
radiative emissions from H$_3^+$ molecular ions which are created by
photoionization of H$_2$ molecules and related ion chemistry \citep{Yelle2004}.

Recently, \citet{Lecavelier2004} and Jaritz et al. (2005) discussed
the Roche lobe effects on hydrodynamic mass loss from the ``Hot
Jupiters'' atmospheres and argued that for some close-in exoplanets,
due to the expected high exospheric temperatures, the exobase level
$r_{\rm exo}$ can reach the Roche lobe $r_{\rm Rl}$ before classical
hydrodynamic blow-off conditions may develop and, thus, affect their
atmospheric loss rates. The Roche lobe is defined as the last
equipotential surface around a planet where its gravitational
potential energy is zero. Below this boundary, the surfaces of
constant potential encompass the host star. Because atmospheric
particles at and beyond the Roche lobe can escape unhampered,
\citet{Lecavelier2004} argued that the Roche lobe can be seen as an
equivalent to the exobase level. However, there is an important
difference between the Roche lobe and the exobase, since at the
Roche lobe and above it a planetary atmosphere is not bound to its
planet by gravitational forces and can freely escape, whereas at the
exobase level (where $r_{\rm exo} < r_{\rm Rl}$) gas temperature
should be above some critical value for blow-off to occur.

In the typical blow-off scenarios considered, for example, by Watson et al.
(1981) and Jaritz et al. (2005) it is assumed that the stellar X-ray and EUV
radiation is absorbed mainly at the so-called expansion radius $r_{1}$ (Watson
et al., 1981) or at $r_{\rm exp}$ (Jaritz et al., 2005) which is located for
most ``Hot Jupiters'' at a planetocentric distance close to $r_{\rm Rl}$ or
above. Therefore, Jaritz et al. (2005) proposed that, depending on a planetary
and stellar mass, planetary size, stellar type and orbital distance, both
scenarios, the classical hydrodynamic blow-off, and the so-called ``geometrical
blow-off'', as expected by Lecavelier des Etangs et al. (2004), may occur.

However, hydrodynamic model simulations by Yelle (2004) which include
photochemistry of a hydrogen-rich atmosphere, indicate that the optical density
for the stellar XUV radiation absorption $\tau_{\rm XUV}$ by evaporating
hydrogen is $\ll 1$ at $r_{\rm Rl}$ or at distances corresponding to Watson's
$r_{1}$, so that the main part of the XUV radiation is absorbed at lower
altitudes. This is also in agreement with a recent study by Kulikov et al.
(2006) who showed that in a terrestrial type planetary atmosphere even at very
high XUV fluxes the bulk of the XUV radiation is absorbed at a distance $r_{\rm
XUV}$ (where $\tau_{\rm XUV} = 1$) which is much closer to the planetary radius
$r_{\rm pl}$ than $r_{1}$. At altitudes which are much higher than $r_{\rm pl}$
or $r_{\rm XUV}$, the absorption optical depth $\tau_{\rm XUV}$ of an
evaporating atmosphere is much less than 1. And  below $r_{\rm XUV}$, where the
optical depth $\tau_{\rm XUV} \gg 1$, in situ XUV atmospheric heating becomes
negligible.

The aim of this paper is to show that for some of the observed
close-in ``Hot Jupiters' the Roche lobe can be fairly close to a
planet and its exobase level, which can substantially enhance the
mass loss of a planetary atmosphere as compared to the classic
blow-off conditions, and may thus have important evolutionary
implications.

\section{Roche lobes of close-in exoplanets}
We consider two spherical masses, $M_{\rm pl}$ (exoplanet) and $M_{\rm star}$
(star), separated by an orbital distance $d$, which rotate around their common
center of mass. In the rotating coordinate system, the energy per unit mass of
a test particle in the ecliptic plane is given by Paczy\'{n}ski (1971)
\begin{equation}
\Phi=-\frac{G M_{\rm pl}}{r_{\rm a}}-\frac{G M_{\rm star}} {r_{\rm
b}}-\frac{G\left(M_{\rm pl}+M_{\rm star}\right)s^2}{2d^3},
\end{equation}
where $s$ is the distance from a given point to the center of
mass, $r_{\rm a}$ and $r_{\rm b}$ are distances to the centers of
the exoplanet and the star, respectively, and $G$ is Newton's
gravitational constant.

The first term in eq. (1) represents the potential of the exoplanet and the
second the potential of the star. The third term is the result of the orbital
motion of the whole system. By introducing dimensionless quantities $\delta =
M_{\rm pl}/M_{\rm star}$, $\lambda = d/r_{\rm pl}$, $\eta = r_{\rm a} /r_{\rm
pl}$, we analyze the variation of the potential along the axis which connects
the exoplanet with its host star
\begin{eqnarray}
\Phi(\eta) &= &\Phi_0 \left[- \frac{1}{\eta} - \frac{1}{\delta (
\lambda- \eta)} \right. \nonumber \\
& & \left.
-\frac{1+\delta}{\delta}\left(\lambda\frac{1}{(1+\delta)} - \eta
\right)^2 \frac{1}{2 \lambda^3} \right ],
\end{eqnarray}
where
\begin{equation}
\Phi_0 = G M_{\rm pl} /r_{\rm pl}.
\end{equation}
For this potential, there are two locations, the Lagrangian point
L1 and the Lagrangian point L2 which are saddle points. Both
points are rather close to each other for a small ratio of mass
\citep{Gu2003}
\begin{equation}
r_{\rm
L1,L2}=\left(\frac{\delta}{3}\right)^{1/3}\left[1\mp\frac{1}{3}\left(\frac{\delta}{3}\right)^{1/3}\right]d,
\end{equation}
therefore, we consider as the Roche lobe boundary $r_{\rm Rl}$,
approximately
\begin{eqnarray}
r_{\rm Rl} \approx \left(\frac{\delta}{3}\right)^{1/3}d.
\end{eqnarray}
The potential difference between the Roche lobe boundary and the
planetary radius (surface) can be written as
\begin{eqnarray}
\Delta\Phi &= &\Phi_0 \frac{(\eta -1)}{\eta}\left[ 1 -
\frac{1}{\delta} \frac{\eta}{\lambda^2}\frac{(\lambda (1+\eta) -
\eta)}{(\lambda - 1)(\lambda-\eta)} \right. \nonumber \\
& & \left.- \frac{(1+\delta) \eta(1+\eta)}{2 \delta
\lambda^3}\right].
\end{eqnarray}
By assuming $d \gg r_{\rm Rl} > r_{\rm pl}$ and $M_{\rm star} \gg
M_{\rm pl}$ what means $\lambda \gg \eta
> 1$ and $\delta \ll 1$, we can simplify the expression for the
potential difference
\begin{equation}
\Delta\Phi = \Phi_0 \frac{(\eta -1)}{\eta}\left[ 1 - \frac{3}{2}
\frac{\eta (1+\eta)}{\delta \lambda^3}\right].
\end{equation}
Introducing the dimensionless quantities $\lambda$ and $\eta$ in eq. (5), we
can further simplify eq. (7) for the potential difference to
\begin{eqnarray}
\Delta\Phi = \Phi_0 \frac{(\eta -1)^2 (2\eta +1)}{2 \eta^3}.
\end{eqnarray}

\section{Implications for atmospheric blow-off}
By considering a hydrodynamic regime of atmospheric escape we
apply the energy conservation equation
\begin{eqnarray}
\Gamma \left[m \Delta\Phi + \frac{m
v^2}{2}+\frac{5}{2}k\left(T_{\rm Rl}-T_0\right)\right] =
\int_{r_{\rm pl}}^{r_{\rm Rl}} q r^2 dr,
\end{eqnarray}
\begin{table*}[t]
\begin{center}
\begin{tabular}{lcccccccc}
\hline
Exoplanet & star type &$M_{\rm \star}$
[$M_{\rm \odot}$]& $M_{\rm pl}$ [$M_{\rm Jup}$]& $r_{\rm pl}$ [$r_{\rm Jup}$]& $r_{\rm Rl}$ [$r_{\rm pl}$]&$1/K$ & K & $d$ [AU]\\
\hline \hline
HD209458b    & G0V & 1.05 & 0.69 & 1.43 & 3.9 & 2.10
& 0.47
& 0.045\\
 OGLE-TR-56b  & G & 1.04 & 1.45 & 1.23 & 3.0 & 2.39 &
0.41 & 0.023\\
 OGLE-TR-132b & F & 1.34 & 1.01 & 1.15 & 3.5 & 1.90
& 0.52 & 0.031\\
 OGLE-TR-113b & K & 0.77 & 1.35 & 1.08 & 3.6 &
1.75 & 0.57 & 0.023 \\
TreS-1       & K0V& 0.87 & 0.75 & 1.08 & 4.9 & 1.48 & 0.67 &
0.023\\
OGLE-TR-111b & G or K & 0.82 & 0.53 & 1.00 & 5.8 & 1.34 & 0.74 & 0.047 \\
OGLE-TR-10 b &G or K  & 1.22& 0.57 & 1.24 & 3.7 & 1.9  & 0.52 & 0.042 \\
\end{tabular}
\end{center}
\caption{Factors $1/K$ and $K$ for 7 ``Hot Jupiters'' where the
planetary mass and radius is known. The planetary and stellar
parameters for HD209458b \citep{Barman2002,Vidal-Madjar2003},
OGLE-TR-56 b \citep{Burrows2004,Baraffe2004}, OGLE-TR-132 b
\citep{Moutou2004}, OGLE-TR-113 b \citep{Bouchy2004}, OGLE-TR-111
b \citep{Pont2004}, OGLE-TR-132 b \citep{Bouchy2004,Moutou2004},
OGLE-TR-10 b \citep{Bouchy2005} and TreS-1 \citep{Alonso2004} were
used for the calculation of $r_{\rm Rl}$ and $K$.}\label{tab:1}
\end{table*}
where $\Gamma$ is the loss rate of particles per steradian, $q=q_{\rm
XUV}-q_{\rm IR}$, with the XUV volume heating rate $q_{\rm XUV}$ and the
cooling rate $q_{IR}$ due to IR emitting molecules like H$_3^+$ (Yelle, 2004),
$m$ is the mass of the evaporating particles, and $v$ is the outflow bulk
velocity at the Roche lobe boundary $r_{\rm Rl}$, $k$ is the Boltzmann
constant, $T_{\rm Rl}$ is the temperature of hydrogen at the Roche lobe $r_{\rm
Rl}$ and $T_0$ is the temperature at a distance close to $r_{\rm pl}$, which is
about the effective radiative temperature $T_{\rm eff}$ of the exoplanet.
Substituting the potential difference $\Delta\Phi$ into (9), we obtain
\begin{eqnarray}
\Gamma =\frac{\int_{r_{\rm pl}}^{r_{\rm Rl}}{q r^2
dr}}{\left[\frac{m M_{\rm pl} G K}{r_{\rm pl}} + \frac{m
v^2}{2}+\frac{5}{2}k\left(T_{\rm Rl}-T_0\right)\right]},
\end{eqnarray}
where the factor $K$ is
\begin{eqnarray}
K =\frac{(\eta-1)^2 (2\eta+1)}{(2\eta^3)} < 1.
\end{eqnarray}
Neglecting the kinetic energy term $mv^2/2$ and also the thermal energy
$5k/2(T_{\rm Rl}-T_0)$ and introducing the planetocentric distance $r_{1}$
\begin{equation}
r_1=\left(\frac{\int_{r_{\rm pl}}^{r_{\rm Rl}}qrdr}{I_{\rm
XUV}}\right)^{\frac{1}{2}},
\end{equation}
we can obtain the energy limited escape rate equation similar to that derived
by Watson et al. (1981), which corresponds to the distance below which the
incoming stellar XUV radiation $I_{\rm XUV}$ is absorbed by an evaporating
atmosphere
\begin{equation}
\Gamma=\frac{r_{\rm pl} r_1^2 I_{\rm XUV}}{ m M_{\rm pl} G K}.
\end{equation}
Fig. 1 illustrates the difference between the XUV absorbtion radius $r_{1}$ of
Watson et al. (1981), the radius $r_{\rm XUV}$ where the bulk XUV flux is
absorbed and the energy absorption function $q_{XUV}$ has its peak, and the
Roche lobe $r_{\rm Rl}$. Depending on the planetary and stellar parameters,
$r_1$ can be located outside the Roche lobe. Due to the low number density of
the evaporating hydrogen at planetary distances around and above $r_{\rm Rl}$,
the optical density of the XUV radiation $\tau_{\rm XUV}$ is $\ll 1$ and the
main part of the stellar XUV radiation is absorbed at $r_{\rm XUV}$ which is
close to $r_{\rm pl}$. This altitude $r_{\rm XUV}$ corresponds generally to the
thermosphere and ionosphere.
\begin{figure}
\centering
\includegraphics[height=4.2cm]{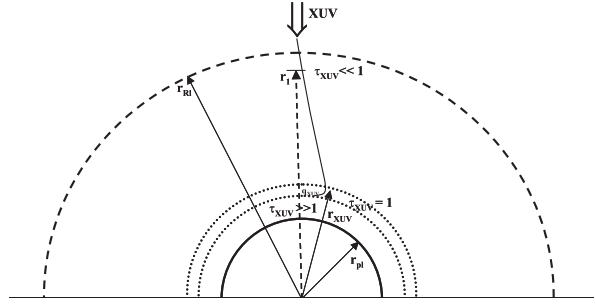}
\caption[]{Illustration which compares $r_{\rm pl}$, $r_{\rm XUV}$,
$r_{\rm Rl}$, $q_{\rm XUV}$ with the XUV absorbtion radius $r_{1}$.
If the exoplanet experiences hydrodynamic conditions, blow-off
occurs at the altitude level $r_{\rm XUV}$ where the XUV energy
absorption function $q_{\rm XUV}$ has its maximum.}
\end{figure}

If the bulk of the stellar XUV radiation is absorbed in the thermosphere at
altitudes $r_{\rm XUV} < r_{\rm Rl}$, the particle loss can be enhanced
substantially by the effect of the Roche lobe which manifests itself in the
loss enhancement factor $1/K$ in eq. (13). Table 1 and Fig. 2 show the Roche
lobe induced mass loss enhancement factor $1/K$ (and also $K$) as a function of
$r_{\rm Rl}$ in planetary radii $r_{\rm pl}$.

One can see from Table 1 that OGLE-TR-56b at an orbital distance of about 0.023
AU experiences the strongest enhancement of the Roche lobe affected evaporation
resulting in a factor of about 2.4, while the Roche lobe induced enhancement of
mass loss at HD209458b at 0.045 AU is about 2. As one can see from Fig. 2, the
mass loss of ``Hot Jupiters'' which orbit at $\leq 0.02$ AU around their host
stars, could be dramatically larger, because $r_{\rm Rl}$ would move closer to
$r_{pl}$ and that enhances the evaporation.
\begin{figure}
\centering
\includegraphics[height=6.0cm]{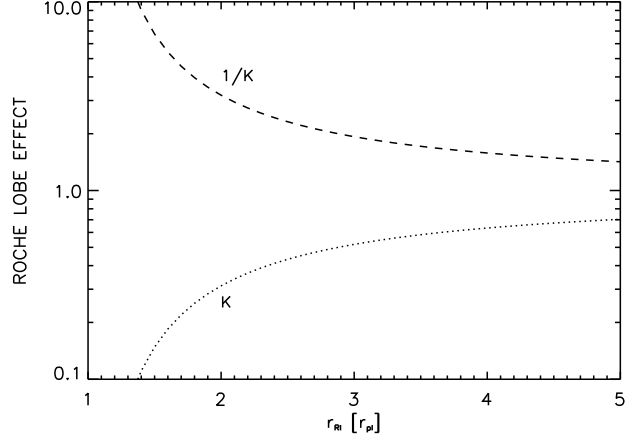}
\caption[]{Roche lobe induced factor $K$ and atmospheric mass loss
enhancement 1/$K$ as a function of $r_{\rm Rl}$ normalized to
$r_{\rm pl}$.}
\end{figure}
One more important effect due to the Roche lobe on ``Hot Jupiters'', which
experience hydrodynamic conditions, manifests itself in the potential energy
difference which is less than the thermal energy at the exobase for blow-off to
occur
\begin{eqnarray}
m\Delta\Phi \leq k T_{\rm max},
\end{eqnarray}
Here $T_{\rm max}$ is the maximum exobase temperature produced by the XUV
heating. This equation yields
\begin{eqnarray}
\frac{Gm M_{\rm pl}K}{k T_{\rm max}} \leq 1.
\end{eqnarray}
One can see that the effect of the Roche lobe helps to satisfy this condition
for lower temperatures than expected for the classic blow-off (that is for
$K=1$). In other words, if the blow-off temperature for an exoplanet without
the effect of the Roche lobe ($K\approx 1$) is, for example, 10000 K, a similar
exoplanet, but which is closer to its host star, may start to evaporate
hydrodynamically due to the Roche lobe effect at about 5000 K if the factor $K$
is $\approx 0.5$ like it is for OGLE-TR-132b (see Table 1).

This result is very important, because it indicates that the effect
of the Roche lobe can enhance the possibility that ``Hot Jupiters''
may reach hydrodynamic blow-off conditions more easily, even if
their atmospheres have a high amount of molecules like H$_3^+$,
which act as IR-coolers in the thermosphere. Both effects, the
enhanced mass loss and the higher probability that ``Hot Jupiters''
reach hydrodynamic blow-off conditions at very close orbital
distances to their host stars, may enhance the evaporation rate. The
results of our study have to be included in the statistical
mass-radius analysis of hot exoplanets expected to be detected
during the CoRoT mission in the near future.

\section{Conclusions}
Our study shows that hydrodynamically driven atmospheric mass loss from ``Hot
Jupiters'' at close orbital distances that are much less than 0.05 AU, may be
strongly enhanced due to the Roche lobe effect as compared to non-affected
exoplanets. We found that the mass loss due to the Roche lobe can be enhanced
several times if the Roche lobe is located closer to a planet at a distance of
a few planetary radii. Furthermore, our study indicates that the Roche lobe
effect may also help ``Hot Jupiters'' to attain hydrodynamic blow-off
conditions even if their exospheric temperatures are lower than those required
for the blow-off to occur in the case of a classic Newtonian gravitational
potential of a planet. Both effects may have a strong impact on the atmospheric
evolution of short periodic hydrogen-rich gas giants.

\begin{acknowledgements}
H. Lammer and H. K. Biernat thank the Austrian Academy of Sciences
``Verwaltungsstelle f\"ur Auslandsbeziehungen''. This work is
supported by the Austrian ``Fonds zur F\"orderung der
wissenschaftlichen Forschung'' (FWF) under project P17100--N08, by
grants \mbox{04--05--64088}, from the Russian Foundation of Basic
Research, and by project No I.2/04 from \"Osterreichischer
Austauschdienst (\"OAD). The authors also thank the Austrian
Ministry for Science, Education and Culture (bm:bwk) and ASA for
funding the CoRoT project. This study was supported by the
International Space Science Institute (ISSI) and carried out in the
frame of the ISSI Team ``Formation, Structure and Evolution of Giant
Planets''.
\end{acknowledgements}

\end{document}